\documentclass[sigconf]{acmart}
\usepackage{amsmath,amssymb,amsfonts}
\usepackage{graphicx}
\usepackage{textcomp}
\usepackage{hyperref}
\usepackage{xspace}
\usepackage{xcolor}

\usepackage{amsthm}

\usepackage{makecell}
\usepackage{ragged2e}  
\usepackage{xspace}
\usepackage{multirow}
\usepackage{booktabs}

\usepackage[ruled,noend,linesnumbered]{algorithm2e}
\usepackage{algpseudocode}
\usepackage{amsthm}
\usepackage[scr=boondox]{mathalfa}
\makeatletter 
\g@addto@macro{\@algocf@init}{\SetKwInOut{executed}{Actor }} 
\makeatother

\newcommand{\sysname}{Libra\xspace}
\theoremstyle{definition}
\newtheorem{definition}{Definition}[section]

\newcommand\blfootnote[1]{%
  \begingroup
  \renewcommand\thefootnote{}\footnote{#1}%
  \addtocounter{footnote}{-1}%
  \endgroup
}

\renewcommand\footnotetextcopyrightpermission[1]{}

\begin{document}

\title{Libra: Fair Order-Matching for Electronic Financial Exchanges}

\author{Vasilios Mavroudis}
\affiliation{University College London}
\email{v.mavroudis@ucl.ac.uk}

\author{Hayden Melton}
\affiliation{Refinitiv}
\email{hayden.melton@refinitiv.com}
\renewcommand{\shortauthors}{Mavroudis and Melton}

\begin{abstract}
While historically, economists have been primarily occupied with analyzing the behaviour of the markets, electronic trading gave rise to a new class of unprecedented problems associated with market fairness, transparency and manipulation. 
These problems stem from technical shortcomings that are not accounted for in the simple conceptual 
models used for theoretical market analysis.
They, thus, call for more pragmatic market design methodologies that consider the various infrastructure complexities
and their potential impact on the market procedures. 

First, we formally define \emph{temporal fairness} and then explain why it is very difficult for order-matching policies to ensure it in \emph{continuous markets}. Subsequently, we introduce a list of system requirements and
evaluate existing ``fair'' market designs in various practical and adversarial scenarios. We conclude that they fail to retain their properties in the presence of infrastructure inefficiencies and sophisticated {\em technical manipulation} attacks. Based on these findings, we then introduce \sysname, a ``fair'' policy that is resilient to gaming and tolerant of technical complications. 
Our security analysis shows that it is significantly more robust than existing designs,
while \sysname's deployment (in a live foreign currency exchange) validated both its considerably low impact on the operation of the market and its ability to reduce speed-based predatory trading.

\end{abstract}

\maketitle
\pagestyle{plain}
\blfootnote{Disclaimer: Views expressed herein do not necessarily reflect those of Refinitiv and do not constitute legal or investment advice.}

\section{Introduction}
The introduction of electronic markets enabled equity, forex and cryptocurrency exchanges around the world to accommodate the trading of various assets worth trillions of dollars between millions of investors. However, this rapid growth gave also rise to concerns about the operational fairness, transparency and information symmetry of exchanges. One of the major points of friction have been the \emph{order matching policies} used by the exchanges and the importance of ``speed''.

The majority of the existing electronic markets rely on a continuous-auction model that matches incoming buy and sell orders in a ``first come, first served'' (FCFS) manner~\cite{budish2015}. Thus, in financial markets---and indeed in many other forms of markets~\cite{brown2016slowing}---there is a clear advantage in being ``fast''. For example, if the quantity of the traded resource is scarce, a participant will need to be fast enough to express their interest before the other investors. Understandably, a slow participant will have a very low chance of being allocated that specific resource. ``Fast'' traders in stock exchanges have a reaction time in the order of a microseconds or perhaps even nanoseconds, while they employ increasingly sophisticated optimization techniques (both in hardware and in software) to achieve marginal improvements~\cite{schneider2012microsecond,ye2013externalities,johnson2012financial,keim2012nanosecond}. 

This phenomenon has been studied extensively and several researchers argue that the existing FCFS market designs incentivize a socially-wasteful technological ``arms race'' in pursuit of speed among competing participants~\cite{farmer2012,mannix2013,harris2013,budish2015,wah2016}. More importantly, speed
advantages are not always the result of smaller reaction times but can be obtained through
\textit{technical manipulation} techniques that exploit flaws, details or inefficiencies in the exchange's
infrastructure~\cite{CMEOptimistic,simonoff2017,Mavroudis19,CME17,CME17}.
For example, the Securities and Exchanges Board of India recently documented several incidents where traders colluded with exchange insiders to unfairly outrace other participants.
The insiders assisted them in receiving faster market updates and order execution by detailing them on the traffic loads of the exchange's servers and by allowing them to connect to faster/less crowded servers intended for contingency.
Additionally, the same traders sought to further increase the information dissemination latency for other market
participants (and thus their reaction time) by actively congesting the servers they were using~\cite{SEBI2109}.
Besides malicious attempts, network congestion, unevenly busy routers/gateways and malfunctioning equipment can all
contribute to uneven delays that affect differently each market participant. 
In fact, there are many examples in the literature and mainstream media where venues have 
been fined for allowing slower participants to ``overtake'' faster ones 
(e.g., \cite{osipovich17,SEBI2109,meltonADSN}).

The problem originates from the gap between economic theory and the unavoidable 
complexities that emerge when theoretically ``fair'' market models (e.g., FCFS matching) are implemented.
While this issue may seem easy to address, the complex and distributed architecture of modern exchanges makes it 
very hard for operators to keep track of all the potential sources of delays and
monitor how they may affect each participant \cite{MacKenzie2019}. Besides monitoring, operators would also need 
to maintain their system perfectly symmetrical and load balanced at all times
with microsecond (or even nanosecond, see \cite{WhiteRabbit}) accuracy.

Instead of trying to eliminate all possible sources of uneven delays,
alternative market designs/order-matching policies have been proposed as solutions~\cite{huang1992,gode2000,farmer2012,harris2013,mannix2013,budish2015,wah2016,budish2017will}. For example, Budish et al.~\cite{budish2015} introduce a policy that treats time as a discrete variable (instead of continuous) and replaces continuous auctions with ``batches''. This allows for multiple orders to arrive at the same (discrete) time and aims to alleviate the importance of (fair or unfair) sub-millisecond differences in submission times. 
It is an open area of research as to the totality of effects of these proposals, with oftentimes contradictory conclusions being described~\cite{duffin2018agent,mizuta2016investigation,brown2016slowing,chakrabarty2018exchange,chen2017value,aoyagi2019strategic,brolley2018order,FIAPTG,donier2016walras,khapko2019speed,tabb2016speed,Tabb2019}. 

In this work, we examine several order-matching policies that have either been proposed in the literature or are currently deployed in electronic exchanges. Towards this goal, we first introduce a generic definition of \emph{temporal fairness} that captures {\em fairness} (i.e., {\em the equal treatment of equals}~\cite{moulin2004fair}) with respect to speed.
Additionally, we outline the required properties that such a policy needs to fulfill in order for it to be 
both fair and practically applicable. Our analysis suggests that existing policies either fail to remain ``fair''
in adversarial settings or introduce delays that are disproportionately large to the discrepancies they try
to address. Furthermore, we discuss some previously undocumented technical attacks and find that
some of the existing policies are prone to them.

Based on these findings, we then introduce \sysname, a carefully thought-out order-matching policy that 
seeks to minimize the processing delay experienced by market participants while ensuring fairness even in the
presence of several infrastructure inefficiencies and sophisticated adversaries. 
Upon introducing the general operation principles of our policy, we consider various matching and manipulation 
scenarios, and verify that \sysname remains fair, fast and does not substantially disrupt existing features of the market that do not pertain to speed. 
Subsequently, we deploy \sysname on a major forex trading venue serving 5000+ users, 900 firms and spanning 50 countries.\footnote{\url{https://www.refinitiv.com/en/products/spot-matching-forwards-matching}} Our results suggest that \sysname, besides ensuring temporal fairness, also contributes to the reduction of predatory practices (e.g., Sniping~\cite{kirilenko2017flash}). Moreover, we verify
that even when deployed with higher latency values, \sysname has only a small impact on
existing trading strategies.

\noindent\textbf{Contributions.}
Overall, this paper makes the following contributions:
\begin{itemize}
    \item Introduces a generic definition of {\em temporal fairness} that constrains all existing viable financial market designs, and crucially one that implementations of those designs {\em can} actually meet.
    \item  Analyzes several market designs and discusses their shortcomings with respect to fairness, resilience to technical manipulation and market impact.
    \item Proposes a novel order-matching policy and evaluates its fairness and effectiveness both analytically and in practice by deploying it in a live forex exchange.
\end{itemize}

\section{Preliminaries}\label{sec:preliminaries}
In this section, we introduce some fundamental concepts of electronic markets and discuss
the operation of modern exchanges.

\subsection{Market Structure}\label{subsec:structure}
\begin{figure}[h]
    \centering
    \includegraphics[width=\columnwidth]{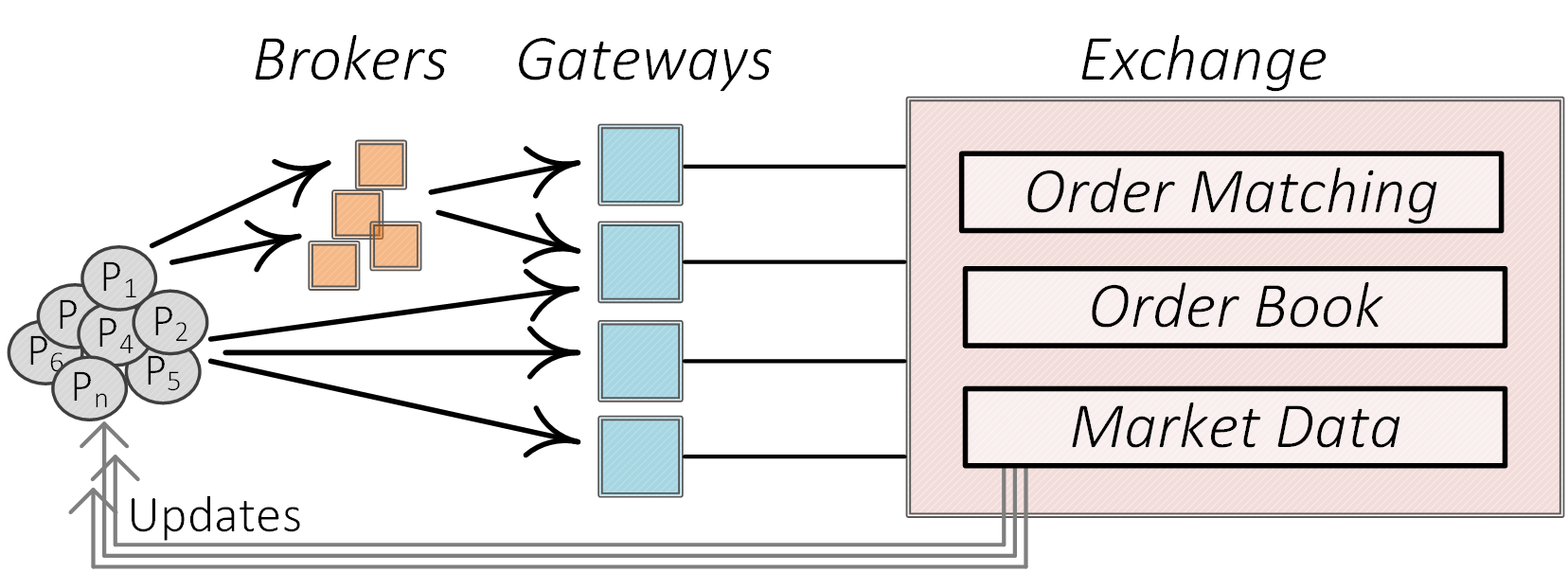}
    \caption{Illustration of the components of an electronic financial exchange. Participants submit their orders
    through their brokers or directly to the exchange's gateways. In
    both cases, the gateways forward the orders to the matching engine that pairs buy with sell orders 
    and updates the order book with the remaining quantities.}
    \label{fig:microstructure}
\end{figure}

Figure~\ref{fig:microstructure} illustrates the different components of an electronic market
and the actors interacting with them. Market participants $P_i$ 
submit their \textit{orders} (e.g., buy stock with symbol `AAPL' at price at most \$100) to 
their brokers who then route them to the market. Alternatively, 
participants who trade larger volumes can submit orders directly to the exchange
(i.e., direct market access). In both cases, the exchange receives the incoming orders 
through its gateways. Due to the
high volume of traffic, exchanges tend to operate several gateways
to balance the load \cite{meltonADSN}. Those gateways, which are located very close to the matching
engine, are tasked with handling incoming orders and forwarding valid ones to it.
Once an order is received by the matching engine, it is placed in the ``book''
where a matching algorithm ranks, pairs and fills it with those of other participants.\footnote{
  A more formal treatment of the limit order book is provided by Gould et al.~\cite{gould2013}.
} Upon a match, 
if its entire open quantity is executed, 
the order is closed and removed from the book. Otherwise, the book's record is updated to reflect only   
the remaining (unfilled) quantity. Finally, the data generation component of the exchange continuously 
updates the participants on the state of both the order book and their pending orders. 

\subsection{Order-Matching Policies \& Competition}\label{subsec:competition}
The extent to which speed affects resource allocation in an electronic financial market depends in large part on the {\em order-matching policy} used. This policy specifies the manner by which a financial exchange purports to process the order messages it receives from market participants.
The predominant order-matching policy in electronic financial exchanges worldwide is {\em price-time priority matching}~\cite{Preis2011}. Under this order-matching policy, the exchange processes order messages (i.e., allocates resources to messages) immediately, in the temporal order in which those messages were received (i.e., in a FCFS manner). 
As a result, ceteris paribus, order received earlier are more  likely to get filled than those received later. 
If the quantity of a resource in a market is scarce, a participant needs to be fast enough to express interest in it before that market {\em clears}. The faster the participant reacts, the higher their allocation chances are. Conversely, if the participant is too slow in expressing their interest,
they may miss the clearing `deadline' and thus have {\em no chance} of being allocated that resource.

In modern exchanges, this speed-related competition has three different manifestations depending on the roles of the competing parties i.e., maker vs. maker, taker vs. taker and maker vs. taker\footnote{In finance literature, ``market makers'' provide liquidity to the market (i.e., post resting orders that populate the limit order book), while ``takers'' remove liquidity by placing orders that are `marketable' (so immediately match with makers' orders  causing those maker orders to be removed from the book). These roles are not fixed and participants transition from maker to taker depending on the side of the trade they take.}. 
In the first case, makers compete with other makers for the best position in the queue at each price level in the limit order book. An order that is earlier in the queue at a price level is much more likely to get filled than one positioned later in the queue~\cite{farmer2012,moallemi2014value}. Similarly, takers compete with other takers for favorably-priced bids and offers of scarce resources~\cite{farmer2012}. The first taker order to the market has access to the full offered volume, while subsequent ones have a lower chance of getting filled, as the volume decreases.
A maker who is trying to cancel their ``stale'' bid or offer may also compete with taker(s) who are trying to lift that bid or offer. This practice is called ``sniping'' and involves fast takers buying assets from or selling assets to slower makers at `stale' bid and offer prices~\cite{budish2015,farmer2012,kirilenko2017flash}.

\subsection{Equal Treatment in Practice}\label{subsec:delays}
In economic theory, fair races for a resource in contention are won by the ``fastest'' participant if the order book operates in a {\em continuous} FCFS  manner~\cite{budish2015,harris2013}. 
However, this assumes that the market's {\em implementation} meets its {\em specification} precisely.
In practice, this type of fairness is hard to meet (or even define) as it is
a tricky and multifaceted concept that always involves a degree of subjectivity~\cite{angel2013fairness,ASX16}.
We now discuss some of the reasons strict-FCFS is hard to realize in practice and why
the fastest participant doesn't always win.\\

\vspace{-0.25em}\noindent\textbf{Jitter.}
Electronic exchanges comprise a plurality of hardware and software components that
exhibit small non-constant variations in their processing times. For example, modern computers are 
optimized for maximum performance (instruction throughput) and thus do not guarantee deterministic execution times. 
This is due to speculative execution, cache eviction in the 
presence of competing processes, charge-to-read and data prefetching that contribute to 
variances\footnote{Other causes of varying latency were identified in \cite{li2014} and include 
thread scheduling policies, NUMA effects, hyperthreading, power saving features, interrupts 
and page faults.} in the process execution runtimes~\cite{Godbolt18}. 
Depending on the component those differences in processing times can vary from
a tenths of nanoseconds to a few hundred microseconds~\cite{Metamako,Godbolt18}.
If the jitter exhibited by an exchange affects each participant to the same degree, then it may not have an appreciable effect on the exchange's fairness.\\

\vspace{-0.25em}\noindent\textbf{Uneven Delays.}
If we use the term `jitter' to describe delays that affect all participants to the same degree, then the other asymmetries 
in the design and the operation of the exchange introduce additional delays that affect different participants differently or unevenly. 
Major modern exchanges have a distributed architecture which allows for better performance
(e.g., parallelization, load balancing) and higher availability (e.g., order gateway redundancy).
However, this also results in discrepancies in the order processing times and asymmetries in the
data dissemination speed, due to the differences in performance among its duplicated components. 
For example, an order gateway used by certain participants may be significantly 
less crowded than others (intentionally~\cite{SEBI2109} or unintentionally) or the 
cables of colocated participants may differ in length thus affecting their latency
 (+/-4 meters or ${\sim20ns}$ per reference  
 ~\cite{eurex16}).
To address this problem, market operators are making an effort to 
equalize the component delays throughout their infrastructure.
While this has reduced the magnitude of the delay discrepancies, 
inter-component differences in delays cannot be fully eliminated 
even if the exchange uses and fine-tunes the exact same hardware and software in
all of its replicated components.\\

\vspace{-0.25em}\noindent\textbf{Technical Market Manipulation.}
This type of manipulation relies on technical means to exploit and extenuate
what might ostensibly seem only to be negligible uneven delays, as well as cause new ones.
Unlike economic manipulation which often involves ``affirmatively misleading'' other participants on matters of supply and demand \cite{levine2015oreos}, 
and is illegal (e.g., spoofing~\cite{lee2013microstructure}), technical manipulation exploits
the implementation details of the exchange itself (e.g., slower vs. faster gateways) and its legal standing 
varies~\cite{ledgerwood2012}.

For example, 
a patent application from a major US-based futures exchange describes an exploit by which participants use 
IP packet fragmentation to gain an advantage over other traders immediately prior to economic events~\cite{CMEOptimistic,Hurd}. 
Given an economic event with two possible outcomes (i.e., a favorable and an unfavorable),
these participants split their orders in several network packets/fragments and submitted all
but one to the exchange. If the outcome of the event was favorable, they submitted also the 
remaining fragment, otherwise they {\em intentionally} corrupted the remaining one (e.g., by invalidating the network/application-layer checksum). 
In another instance, the same exchange found participants sending messages in ``triplicate'' (sending one copy to each of the three gateways each participant was allowed to connect to), so as to improve their chances of outracing their competition by essentially betting one of their gateways was less loaded than the other two~\cite{CME14}.

Participants have also been found to actively influence the operation of the exchange's subsystems 
to create new manipulation opportunities for themselves. The National Stock Exchange of India has been
affected by malicious traders who colluded with exchange insiders to gain unfair advantages over other
participants. Their modus operandi included: (1) getting information about and access to less congested gateways, (2) insider knowledge about the exact opening times of servers as market updates were broadcasted in the
same order participants logged in, (3)  delay competing participants by actively congesting other servers~\cite{SEBI2109}.

Overall, regardless of the legal implications of technical market manipulation\footnote{
Hardware, software and networking optimizations (e.g., TCP/IP tuning for latency over bandwidth) are not necessarily manipulative.}, such practices---seemingly disadvantageously, and in a manner reminiscent of the `arms race' in pursuit of speed ~\cite{budish2015,harris2013}---always result
in unequal treatment of equally fast participants and thus impact the trustworthiness of the market.\\

\subsection{Timescales}\label{subsec:timescales}
Market participants always seek to reduce their reaction and submission time
so as to outrace others competing for the same resources.
One of the most common ways of reducing latency is for trading firms to house their systems in 
the same datacenter with the exchange's matching engine (i.e., colocation). 
This practice may result in a ``100-200 millisecond" advantage over non-colocated participants, and particularly those that are geographically quite distant from the exchange~\cite{arnuk2009latency}.\footnote{
  One response to neutralizing latency differences among geographical dispersed participants involves timestamping the sending times of their messages, and accumulating those messages after their receipt so as to process them by sending time, and not by receipt time~\cite{friedman2018double,glodjo2007global,CBOT1990,Faris2000}. We are only aware of one such exchange that historically may have done this~\cite[p.18]{united1990trading}. 
} 
Nonetheless, the competition between colocated participants remains fierce, as exemplified by the National Stock Exchange of India preferential access colocation scandal \cite{SEBI2109}.

Market participants have now the expertise and the technical capabilities to leverage
speed advantages that last only a fraction of a microsecond or even a few hundred nanoseconds~\cite{schneider2012microsecond,ye2013externalities,johnson2012financial,keim2012nanosecond}.
For example, a public disciplinary notice from a major US-based future exchange reports that one participant sought to gain a speed advantage
by dropping a single character (i.e., 2 bytes) from their order submission messages despite 
the fact their in-house network link had a bandwidth of 10Gb/s~\cite{CME17}.
Interestingly, ``speed advantages'' are now in the same timescale
as the jitter exhibited by processing and networking systems.
Thus, sophisticated market participants can exploit (ostensibly) minuscule latency
discrepancies between different components to gain an advantage over their competitors. 

Due to the complexity and heterogeneity of the exchanges' architectures and infrastructure it is
hard to precisely determine a generic upper bound for the maximum delay discrepancy between
the fastest and the slowest participant (across all exchanges).
For instance, the Metamako switch indicates a jitter of only a few nanoseconds~\cite{Metamako}, while
the variability in the cache/memory fetch time can result in inter-participant latency discrepancies of
several hundred nanoseconds~\cite{Godbolt18}. Similarly, congested routers may introduce discrepancies 
ranging from a few tenths to several hundred microseconds.
In the rest of this paper, we, conservatively and without loss of generality, assume that
the delay discrepancies experienced between colocated participants of an exchange do not exceed 1 millisecond. 
This is in accordance with that reported by a major European exchange that estimates discrepancies to be as high as hundreds of microseconds~\cite[p.16]{eurex16}. 
\section{Temporal Fairness}\label{sec:fairness}
Fairness in financial markets and its relation to ``speed'' has been a topic of much recent interest from academics~\cite{daian2019flash,budish2015,harris2013,budish2017will,wah2016}, practitioners~\cite{meltonSIGecom,meltonADSN,meltonCSCI,melton2018fairware} and even the general public~\cite{lewis2014}.
While most recent works focus on its relation to speed and social welfare, we study it in the context of fairness.
This section defines, in its most general form, the manifestation of fairness in which we are interested and examines the reasons
modern exchanges often fail to meet it.

Two market participants $P_1$ and $P_2$ are said to {\em compete} 
for a trading opportunity $op$ on an exchange $E$ if  responsive to the same  
economic stimulus and they both submit order messages that seek to capture $op$.
A market participant's (P's) {\em speed} is defined to be the time that elapses between 
the receipt of the stimulus and the submission of the order by P\footnote{The discussion on the precise conditions that must be met for the stimulus to have been considered ``received'' and for the order messages to have been considered ``sent'' is left for future work.}. 
In the rest of this paper, we adopt the position of~\cite{budish2015,harris2013} and assume that a market participant's speed is a function of their (direct or indirect) investment in technology i.e., the more a participant spends the faster they are expected to be. We now provide a formal definition of fairness with respect to time.

\begin{figure}
    \centering
    \includegraphics[width=0.5\columnwidth]{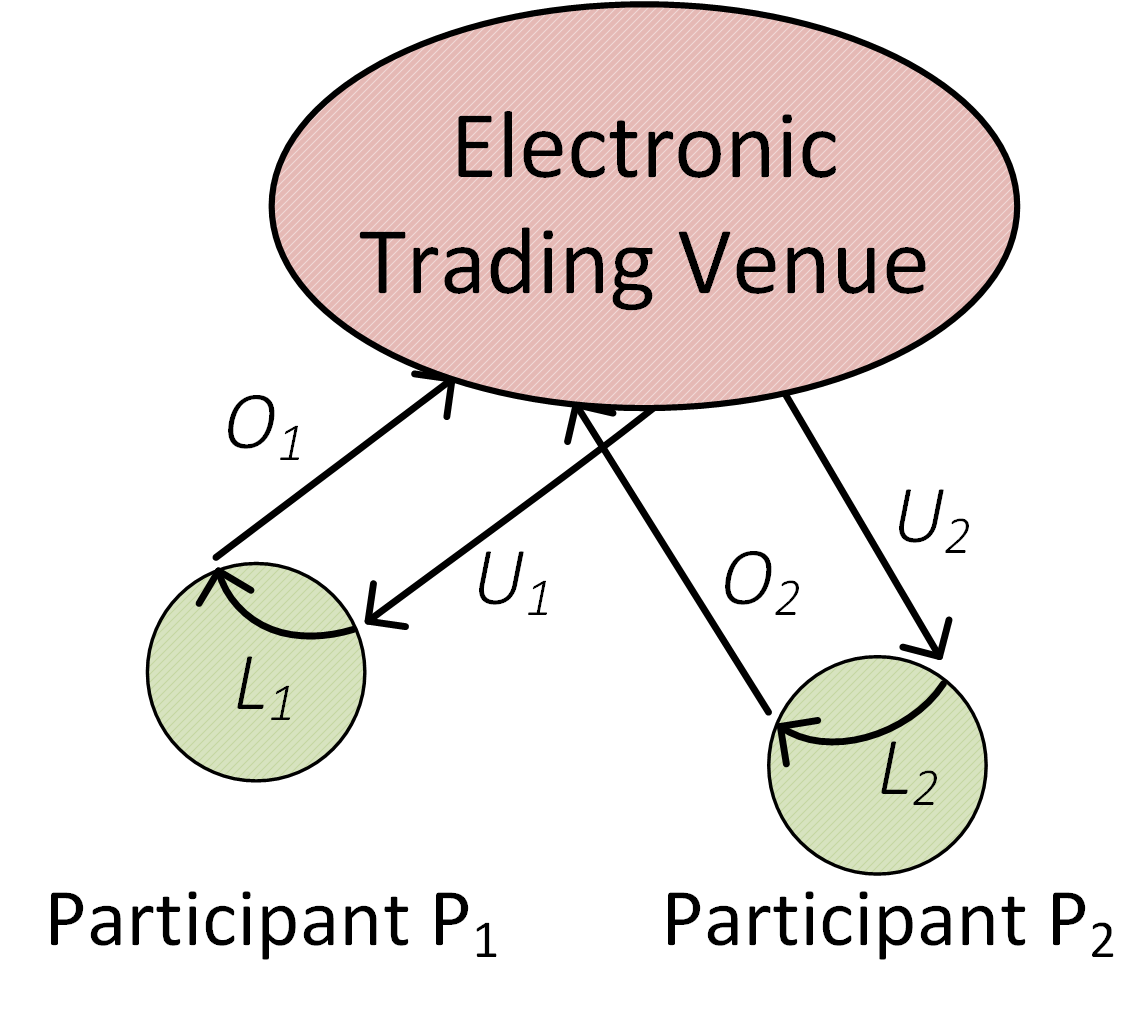}
    \caption{Two market participants $P_1$ and $P_2$ compete for a trading opportunity by sending order messages to an electronic financial exchange $E$ after processing an event update sent by $E$. $U_i$, $L_i$ and $O_i$ denote the update, reaction and transmission latency respectively.}
    \label{fig:generic}
\end{figure}

\theoremstyle{definition}
\begin{definition}{Temporal Fairness:}\label{def:fairness}
A financial exchange $E$ is considered temporally fair if among all pairs of participants on it  
the probability that a slower participant succeeds in capturing a trading opportunity at the expense
of a faster participant (who, responsive to the same stimulus, is also competing for the same opportunity) does not exceed 0.5.
\end{definition}

While we cannot prove, per se, that this general definition of temporal fairness is ``correct''---rather we must ultimately accept it as axiomatic---we at least seek to show it is
not inconsistent with specific notions of fairness and speed that have appeared in the literature. 
Note also that, besides ensuring that slower participants cannot be (statistically) advantaged over faster ones,
this definition allows for market designs/order-matching policies that seek to bring slower and faster participants onto ``equal footing''.

\subsection{Points of Unfairness}\label{subsec:points}
While ``temporal fairness'' may seem obvious and trivial to achieve, there are many examples in the 
literature and mainstream media where venues have been fined for allowing slower participants to consistently ``overtake'' 
faster ones.
For example, the New York Stock Exchange (NYSE) was fined for publishing market data earlier to
a set of participants~\cite{osipovich17}, while the National Stock Exchange (NSE) of India was found to provide 
``preferential access'' to certain customers~\cite{SEBI2109}. Further incidents of real-life unfairness
in electronic exchanges are enumerated in~\cite{meltonADSN}.
We now take a more in-depth look into the different types of delays
that may result in fairness violations.

\begin{figure}
    \centering
    \includegraphics[width=0.7\columnwidth]{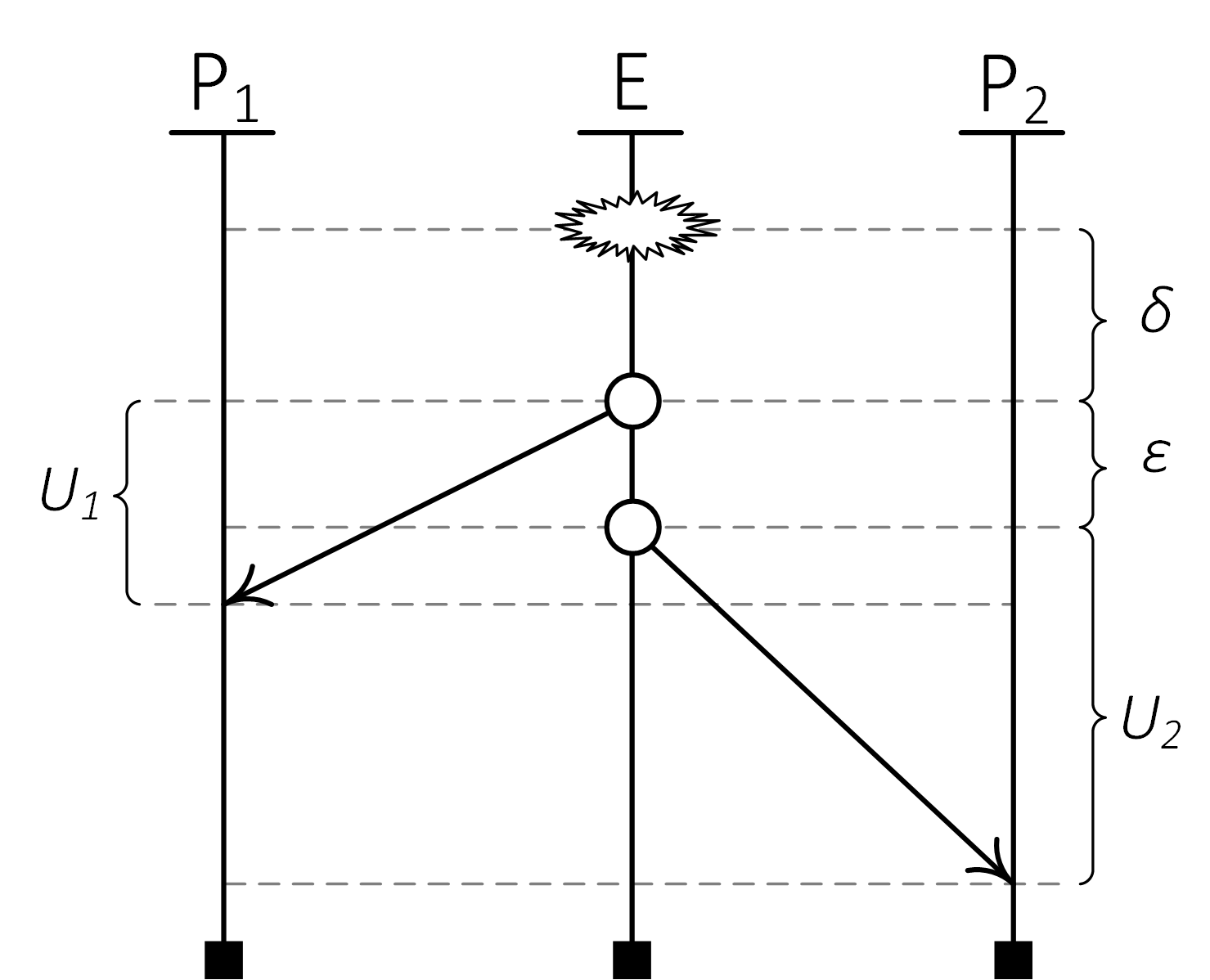}
    \caption{An exchange $E$ sends updates to two market participants $P_1$ and $P_2$. $\delta$ denotes the time it elapses from an event occurring before the exchange sends out the first update to $P_1$, while $\epsilon$ is the time it elapsed until the update to $P_2$ was sent, after it was sent to $P_1$.}
    \label{fig:updates}
\end{figure}

Given an exchange $E$ and two market participants $P_1$ and $P_2$, 
we detail the precise conditions under which the slower (in reaction time)
participant ($P_1$) may be advantaged over $P_2$ that reacts more quickly to stimuli.
Figure~\ref{fig:updates} illustrates the exchange $E$ pushing an event update
to both participants. The total update latency experienced by the individual participants 
is the sum of the individual delays introduced during this process. 
The first cause of delay is the time $\delta$ that the exchange takes to 
prepare the update to be sent to traders. This work focuses on intra-venue trading 
and hence $\delta$ does not impact the fairness of the exchange, as it affects all the 
participants equally. However, in scenarios where participants trade in several
exchanges (i.e., inter-venue trading), a slow participant trading in $E_1$ with a low 
$\delta$, may be advantaged over a faster participant trading in $E_2$ with a higher 
$\delta$ (see e.g., \cite{menkveld2017need,brolley2018order}).

Unlike $\delta$ who affects all the participants equally, $\epsilon$ is the time that elapses 
from when $E$ sends an update to $P_1$ until the same update is sent to $P_2$.
Such delay discrepancies result in substantial differences in the update latency 
experienced by the participants. For example, large $\epsilon$
discrepancies are observed in systems who \textit{unicast} their updates 
to each participant sequentially (see also Section~\ref{sec:existing}) \cite{SEBI2109,meltonFairCredit,howorka2010distribution}.
Besides these, updates may be further delayed by other factors that 
stand between the exchange and the participants. 
For example, margin-of-error differences in ostensibly equal length network links and network switch 
congestion are two common causes of such delays.
The total delay
discrepancy introduced by these factors is given by $U_2-U_1$.

Upon receiving an update participants react to the stimulus by sending their orders to the exchange.
We assume that $P_1$ and $P_2$ receive the event update simultaneously. However, $P_1$ has a slower
reaction time than $P_2$ and thus submits its order with a delay $\zeta$ after $P_2$. Normally, 
this would mean that $P_2$'s order would appear in the limit order book earlier than $P_1$'s.
In practice, there are several factors 
that could delay $P_2$'s order, thus allowing $P_1$ to unfairly overtake it before reaching
the exchange's order gateways $Gw_1$ and $Gw_2$. These are usually factors that the
exchange is not in control of e.g., different network cable lengths, busy network links, network serialization, etc. 

For simplicity, we assume that the two orders arrive simultaneously at their respective gateways.
Gateways and order handling systems within the exchange are another source of uneven delays ($\theta$),
as even if their capacity and specifications are identical, several other factors could impact their
performance (see also Section~\ref{subsec:delays}). For example, a third market participant may be 
using one of the gateways to submit orders at an effective rate of millions per second\footnote{
  It is unlikely a participant could literally submit millions of messages per second because of throttling that most exchanges implement as a risk mitigation measure at their gateways. It is not necessary however to congest the gateway for a full second; even congestion causing momentary delays of nano- or micro-seconds from a burst of messages with that {\em effective rate} can be enough to disadvantage other participants in a FCFS market.
} thus congesting it severely. 

\begin{figure}
    \centering
    \includegraphics[width=0.8\columnwidth]{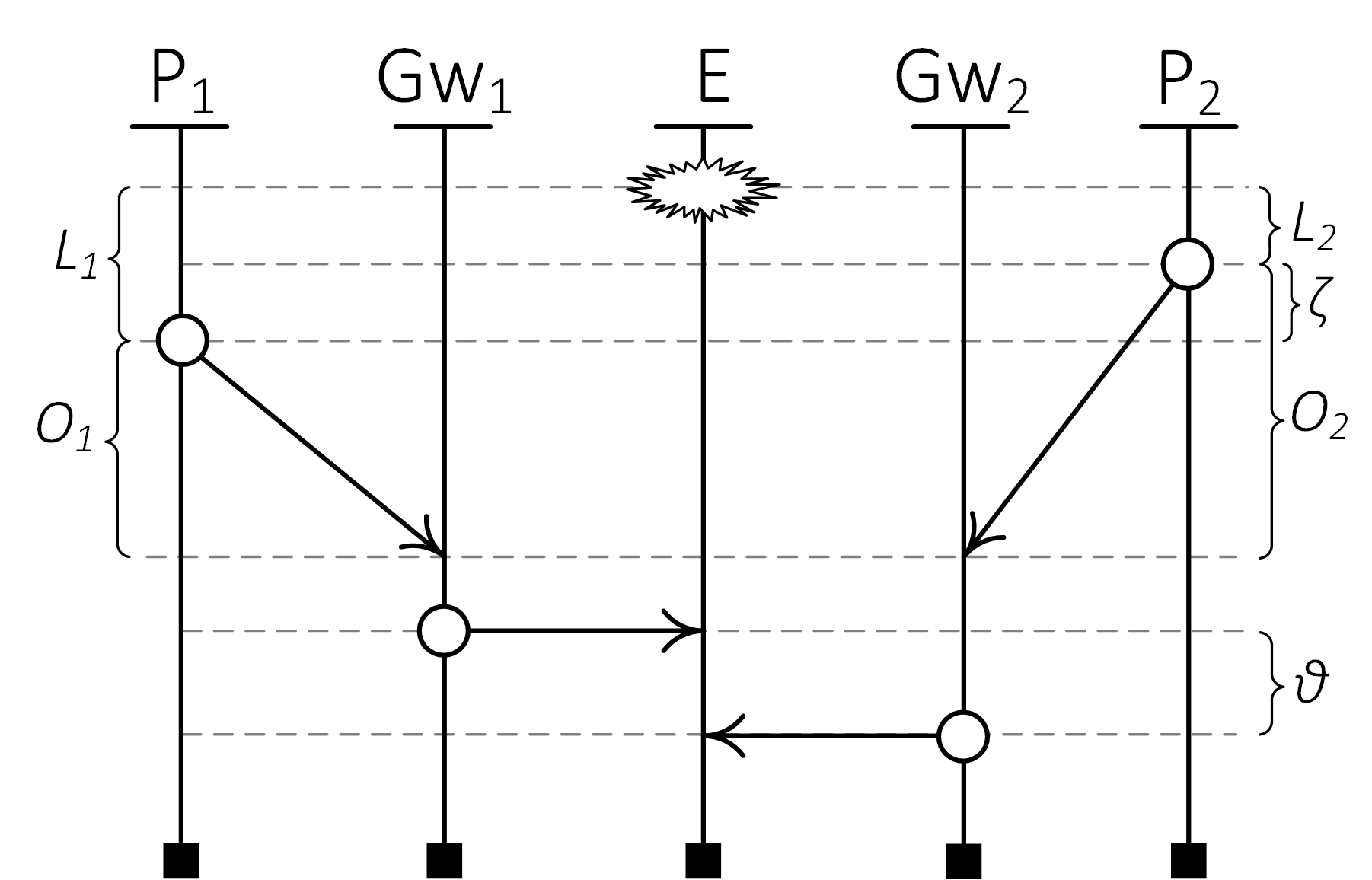}
    \caption{Two market participants $P_1$ and $P_2$ submit their orders to an exchange $E$ through the exchange's order gateways $Gw_1$ and $Gw_2$. $L_1$ and $L_2$ denote the response time of each participant to the common stimulus, while $\zeta$ is the extra time it takes for $P_2$'s order to reach the exchange (from the time it was sent) compared to $P_1$'s order. $\theta$ denotes the latency difference between the gateways $Gw_2$ and $Gw_1$.}
    \label{fig:orders}
\end{figure}

From the above we conclude that there are several points in both the market data dissemination
procedure and the order submission one that can introduce delays affecting participants unevenly.
In the following section, we explain why those points of unfairness cannot be always mitigated.

\subsection{Practical Considerations}\label{sec:practicalconsiderations}
While initially the above latency problems may seem easy to address, in practice
there are several technical constraints that make it very hard for exchange operators
to completely dissipate ``unevenness''. For this reason, we argue that the matching 
processes of an exchange must be designed in such a manner that fairness is preserved
even if its implementation unavoidably introduces uneven delays. Put another way, and has been observed by Roth, we ``need to deal with all of a market's complications, not just its principal features" \cite{roth2002economist}.\\

\noindent\textbf{Data Distribution Latency.}
As seen in Figure~\ref{fig:updates}, exchanges continuously send update messages to their
participants providing information on the contemporaneous state of their limit order book.
Ideally, these updates would arrive to all the participants simultaneously. However, in practice 
the arrival times of those messages can differ by up to 1 millisecond (see also Section~\ref{subsec:timescales}).
In Section~\ref{subsec:points}, we identified the two types of delays that can affect the update latency experienced by intra-venue participants: $\epsilon$ and $U_2-U_1$. $\epsilon$ is the time that elapses from 
when an update is sent to $P_1$ until the same update is sent to $P_2$.
While this can occur for several reasons (e.g., non-optimal exchange processes), we are mostly interested in
causes that persist even in exchanges who have optimized their processes to prevent uneven delays.
For example, discrepancies in the arrival times of updates may occur solely due to the 
protocols realized by the routers of the exchange. If the exchange is using
unicast transmission to send its updates, then by definition the TCP/UDP packets will be sent
sequentially and not simultaneously. Instead the exchange must use one-shot multicast to ensure
that all the updates are sent at the same time\footnote{Multicast based on fanout-spitting or application-layer multicast overlay services suffer from delay problems similar to those of unicast~\cite{prabhakar1997multicast}.}. However, even in this case the inter-arrival delay of the packets is reduced but not eliminated~\cite{klocking2001reducing}.
$U_2-U_1$ is due to margin-of-errors differences in cable lengths, congested
network routers or severed network links and can result in certain participants being systematically
disadvantaged over other slower-to-react participants.\\

\noindent\textbf{Order Submission Latency.}
Latency discrepancies in order submissions occur due to $\zeta$ and $\theta$ delays. $\zeta$ is similar to 
$U_2-U_1$ and is attributed to network links, routers and other network conditions.
$\theta$ is due to inefficiencies and complexities in the exchange's infrastructure.
For example, $\theta$ can be the result of a slower or congested order gateway.
Such discrepancies are common in systems that have a non-monolithic architecture that (via {\em horizontal scaling}) distributes the load 
between several replicated components. In most non-trading use cases, discrepancies in the order of milliseconds
do not have any impact on the usability or the operation of the system.
However, as discussed in Section~\ref{subsec:timescales} trading is one of the few applications
that can be severely affected even by seemingly negligible differences.

\section{Model \& Assumptions}
As discussed in Sections~\ref{sec:preliminaries} and~\ref{sec:fairness},
dissipating uneven delays in an exchange is a best-effort process 
that minimizes but does not fully eliminate discrepancies.
The causes of uneven delays often fall outside of the operator's
control and many of them are due to technological factors that are very hard
to account for and balance in high-load, distributed systems. 
However, an exchange can measure with reasonable accuracy the error-margins
of its systems and subsequently estimate the maximum latency discrepancy that
an unlucky participant may be exposed to (see also Section~\ref{subsec:timescales}).
Based on this knowledge, it can then seek to ``equalize'' all the orders that arrive
within specific time-margins \cite{meltonCSCI,melton2015Measure}. Note that this is compatible with our definition of 
fairness (Section~\ref{sec:fairness}) which allows for slower participants to be
brought on ``equal footing'' with faster ones.
We now outline how such a policy fits into the general
order processing pipeline, the threat model it should be secure under and the
properties it should preserve.

\subsection{System Model}\label{subsec:sysmodel}
\begin{figure}
    \centering
    \includegraphics[width=\columnwidth]{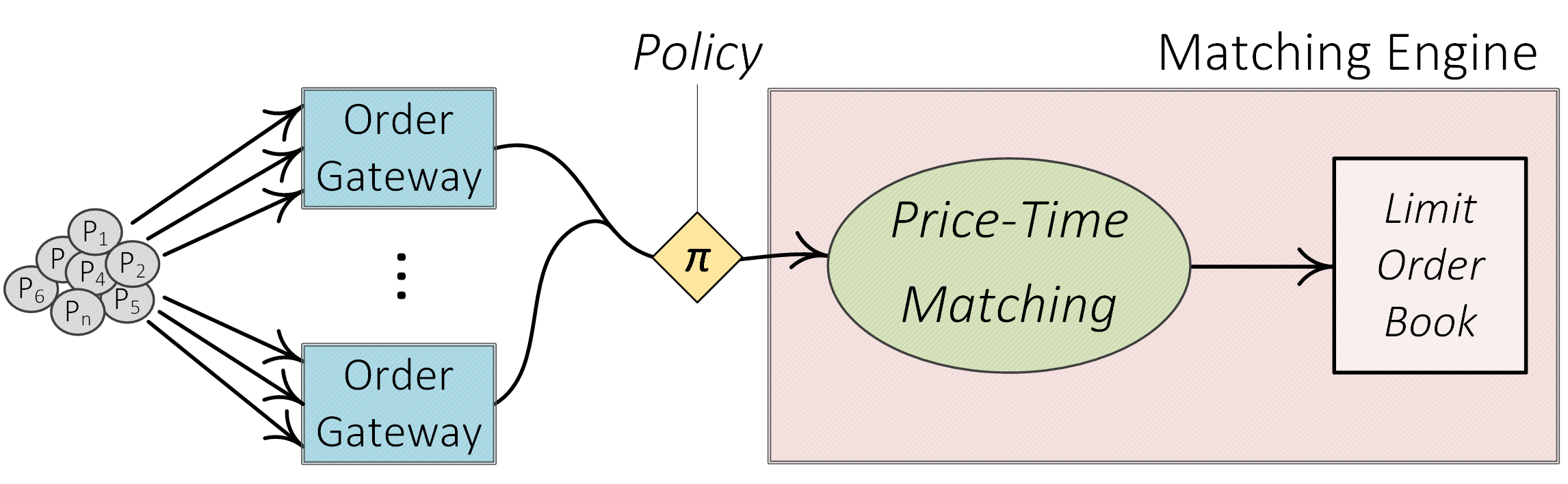}
    \caption{Reordering policies are realized by components that ``intercept'' all order messages before they reach the matching engine. Depending on the magnitude of their inter-arrival time, competing orders are assigned equal or different execution priorities.}
    \label{fig:arch}
\end{figure}

Order ``equalization'' is enforced 
through a \textit{reordering policy} that assigns equal priority to all the 
orders that arrive within a certain time span. Such policies are realised by a component 
that ``intercepts'' all incoming orders before they reach the matching
engine (Figure~\ref{fig:arch}).
For each intercepted order message, the component executes a reordering algorithm 
which determines its priority in relation to the orders that precede it
and those that it precedes. Once the new ordering of the messages has been established, 
the component forwards them to the matching engine and the limit order book.
While each race for a resource can have only one winner, reordering policies
aim to stochastically equalize the winning probabilities of two ``comparably'' fast 
participants over many races.

\subsection{Threat Model}\label{subsec:threatmodel}
We assume that market participants are rational players that will use all the 
means at their disposal to gain a competitive edge.
However, they will not engage in clearly illegal activities such as spoofing. 
Participants are 
also assumed to (1) have full knowledge of the exchange's design, infrastructure
and processes, and (2) be able to exploit advantages that last only 
a fraction of a microsecond. They are also aware of the transmission latency of their own orders
and can modify the structure of their packets, send redundant copies or 
even withhold parts of them. In case the exchange uses a non-FCFS matching policy,
participants are also assumed to have full knowledge of any ``equalization'' policies
and the system's settings (e.g., the duration and starting times of the batches). 

As discussed in Sections~\ref{sec:preliminaries} and \ref{sec:fairness},
the exchange's components introduce both jitter and other uneven delays 
to the latency experienced by market participants (e.g., +/-4 meter variations in the 
cable lengths of colocated participants~\cite[p.6]{eurex16}).
Participants may also collude with exchange insiders who can provide them with 
all these information on uneven delays exhibited by the exchange's components.
However, exchange insiders cannot modify the operation of the market
 (e.g., edit the source code of the matching algorithm).
Exchanges are assumed to have full knowledge of their customers' identities (i.e., Know-Your-Customer) and
the means to identify Sybil accounts~\cite{douceur2002sybil}.

\subsection{Properties}\label{subsec:properties}
An order-matching policy needs to have certain properties
so that it 1) guarantees temporal fairness under our threat model,
and 2) is practically applicable.

\noindent\textbf{Correctness.}
As discussed in Section~\ref{sec:fairness}, an electronic trading venue is temporally fair if
slower participants are never advantaged over a faster ones.
Order-matching policies must always handle orders in a fair manner,
at all possible valid states of the limit order book and for all combinations of order types.

\noindent\textbf{Robustness.} 
Besides correctness, the order-matching policy should be resilient to both system errors and 
gaming attempts. Thus, it must be able to withstand networking issues e.g., severed links,
congested routers and unresponsive servers. Moreover, it should maintain its fairness properties
even at the presence of adversarial, strategic participants (e.g., technical manipulation attacks).

\noindent\textbf{Minimum Impact.} 
While in theory order matching policies can be arbitrarily complex, 
drastic changes in the market structure impact the trading activity negatively.
For example, exchange operators compete fiercely for their market share and
are often wary of long delays that may cause participants to move to other exchanges with 
shorter delays. Thus, designers must strive for minimal divergences from the established market structure,
as every change carries a risk of invalidating the existing trading strategies which in 
turn results in losses of volume and brokerage (e.g., by disenfranchising market makers, or by `breaking' existing profitable trading strategies) \cite{meltonSIGecom,phelps2010evolutionary}.
Moreover, policies should be as simple as possible, so that implementation errors are 
less probable, debugging is easier and the exposed attack surface is minimized (i.e., economy of mechanism).
For example, short (and constant) delays are preferable as they do not conceal network issues and are less likely to
drive participants to route their order-messages to competing venues with faster order execution times~\cite{donier2016walras}, and are less likely to conflate `weak' early trading signals with `strong' later ones \cite{baldauf2018fast}.

\section{Existing Reordering Policies}\label{sec:existing}
We now study various proposals and examine their properties under our threat model.
In all our comparisons, we assume two participants
$P_1$ and $P_2$ that have equal reaction times. However, $P_2$'s updates and 
submitted orders are affected by uneven delays which give an unfair advantage ($d\tau$) to 
$P_1$. Based on the timings reported in Section~\ref{subsec:timescales}, we assume that
the uneven delays experienced by $P_1$ and $P_2$ do not exceed 1 millisecond.
This is without loss of generality and our analysis also applies to exchanges with 
longer or shorter latency error margins or exchanges who want to be fair for
fast but non-colocated participants too.

\subsection{First Come, First Served (FCFS)}
Most electronic financial exchanges implement a market design that is known as the Continuous Limit Order 
Book (CLOB)~\cite{budish2015,gould2013}. This design specifies that order messages are processed by the exchange in the temporal order that they are received. In theory, FCFS guarantees that the fastest participant competing for a trading opportunity will always succeed in capturing it. This implies that the probability of a slower participant capturing a trading opportunity when in competition with a faster participant will be less than 0.5 (it will be 0 with high probability), which satisfies the \textit{correctness} property as it is defined in Section~\ref{subsec:properties}.

\begin{figure}
    \centering
    \includegraphics[width=0.90\columnwidth]{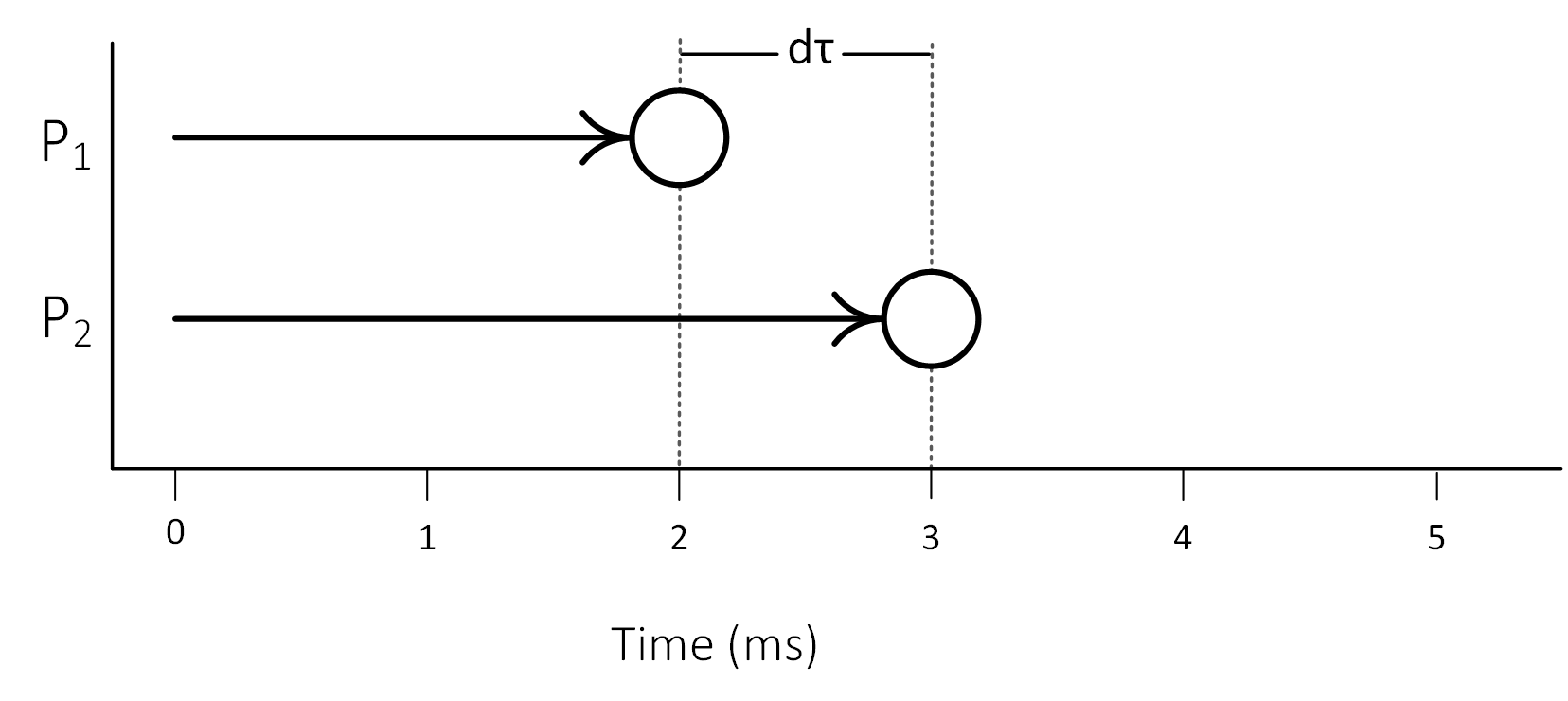}
    \caption{FCFS guarantees that an exchange remains ``fair'' in non-adversarial, perfect conditions. However, 
    if a market participant $P_2$ experiences uneven delays in their updates and/or order submission times (e.g., due to a congested order gateway), FCFS fails to preserve fairness and allows other participants ($P_1$) to exploit that delay to outrace them.}
    \label{fig:fifo}
\end{figure}

However, in practice, FCFS-based systems suffer from robustness issues in the presence of
uneven delays. As seen in Figure~\ref{fig:fifo}, $P_1$ will always outrace $P_2$ despite the 
fact that they both submit their orders at time $t=0$. Thus, FCFS does not fulfill our
\textit{robustness} requirement.

\subsection{Constant Delays}
This policy specifies that an incoming order received by the exchange's engine is delayed by a
fixed amount of time before it is included in the order book. This mechanism is used (or has been proposed) by certain
exchanges~\cite{brown2015adverse,Mavroudis19,icespeedbump2019,lewis2014,Nasdaq2012} and can prevent certain types of unfairness if applied asymmetrically between the participants. More specifically, if the delays are applied only on takers' orders,
then it protects makers from sniping attacks~\cite{kirilenko2017flash,budish2015} that exploit
uneven delays. Unfortunately, this mechanism is only effective when applied asymmetrically and thus
cannot guarantee fairness for all market participants. For example, it cannot prevent unfairness that manifests in 
maker vs. maker races, and in taker vs. taker races.

\subsection{Random Delays}\label{subsec:randomdelays}
Harris~\cite{harris2013} made one of the first proposals that sought to de-emphasize the importance of speed.
Random delays aim to make it equally likely for participants with similar response times to have success in both price-making and price-taking. This design has been implemented by at least one real-world financial exchange~\cite{kellerman2019} and imposes random delays on order messages after they are initially received by the exchange. 
Thus, orders can be matched only after they have been subject to the random (uniformly distributed) delay. 
This policy is simple to specify and implement, and can improve the fairness properties of an 
exchange, as it guarantees that the advantaged participant ($P_1$) won't always be the one that wins.
However, it fails to substantially reduce the advantage of $P_1$ unless
very ``long delays" (relative to differences in speed it seeks to equalize) are used \cite{meltonADSN}.

\begin{figure}
    \centering
    \includegraphics[width=0.9\columnwidth]{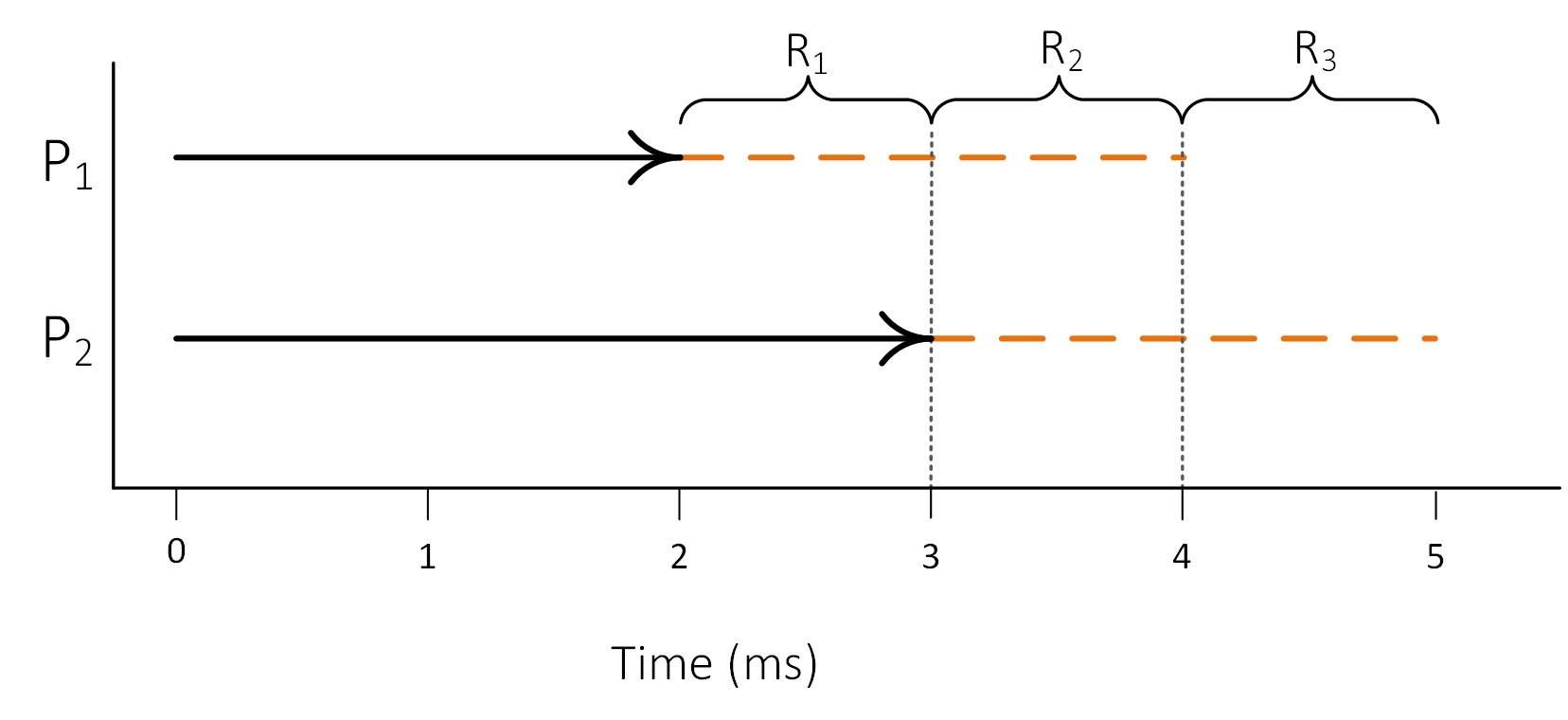}
    \caption{Random delays aim to de-emphasize small time differences in the response times of market participants ($P_1$ and $P_2$). Each incoming order is delayed by a randomly samples period of time before it }
    \label{fig:random}
\end{figure}

In the example of Figure~\ref{fig:random}, the two participants $P_1$ and $P_2$ have a difference 
in the delays they are exposed to $d\tau=1ms$. If we apply a random delay of $2ms$, then the 
orders of $P_1$ and $P_2$ will be processed within the time intervals [2,4] and [3,5] respectively.
If the sampled delay applied on $P_1$'s order is less than $d\tau=1ms$, then $P_1$ always wins the race (Region R1).
Similarly, if $P_2$'s order is delayed for more than $2ms-d\tau=1ms$, then $P_1$ is guaranteed to win (Region R3).
However, $P_2$ will win in 50\% of the races where both orders are executed in Region R2.
The two delays are sampled independently from $Unif[0,2]$ and thus we can easily estimate how likely 
that to happen:
\begin{align*}
Pr(P_1 \in R_2 \wedge  P_2 \in R_2) = \\
Pr(P_1 \in R_2) Pr(P_2 \in R_2) = \\
1/2 * 1/2 =\\
0.25
\end{align*}

We observe that even if we apply a delay that has twice the duration of the uneven delays
present in the exchange, we fail to eliminate unfair advantages and $P_1$ still 
wins $75\%$ of the time. This limits the applicability of random delays as 
there is a direct trade-off between the fairness achieved and the divergence from 
the ``normal'' operation of the exchange. While this may not seem particular worrisome 
in this case, increases in latency exacerbate adverse selection of ``stale quotes"
and create sniping opportunities even against fast participants.
Random delays are also prone to gaming as participants 
may send multiple redundant copies of the same order to increase their
chances of getting a very low latency\footnote{Note that this does not have to do with the participants trying to minimize their 
order submission delay by trying different routes and gateways
(which is already accounted for in the 1ms advantage of $P_1$).} (i.e., multiple lottery tickets). 

\subsection{Frequent Batch Auctions}
An alternative design that has received much 
attention from the academic community---and indeed 
seeks to mitigate the technology `arms race' among market 
participants in pursuit of speed---is known as both the  
{\em frequent call market}~\cite{wah2016} or
{\em frequent batch auction (FBA)}~\cite{budish2015}. 
FBA divides the trading day into fixed-length ($L$) batching intervals, 
and after it accumulates all the orders that arrived within each
interval for joint processing, it matches all the received
``buys'' and ``sells'' at a single clearing price time.

If we assume that the economic events that trigger races between
participants are uniformly distributed with respect to time, then
the probability a participant $P_1$ and a participant $P_2$
(who is by $d\tau$ units of time disadvantaged compared to
$P_1$)\footnote{
  Note that by definition $d\tau > 0$.
}
end up in the same batch is:
$$\frac{L-d\tau}{L}$$ for $d\tau \leq L$. If 
both $P_1$ and $P_2$ end up in the 
same batch then the probability that $P_2$ 
succeeds at capturing the trading opportunity is 0.5, 
which implies that the total likelihood that 
$P_2$ captures the trading opportunity 
is $\frac{1}{2} \cdot \frac{L-d\tau}{L}$.
Thus, a batch interval $L=2ms$
would result in $Pr(P_2\ wins) = 1/2 * (2-1)/2 = 0.25$.
By increasing $L$ to $3ms$, we further increase $P_2$'s winning probabilities
to ${\sim}$33\%. 
Thus, FBA still requires long delays to address small differentials in speed,
while the improvement in fairness diminishes rather fast as the length of the batching intervals increases.
FBA is not prone to the same attacks that affect random-delays but
participants may still be incentivized to compete on speed to ``bang the close'' by submitting their orders
just before the end of a batching period~\cite{farmer2012,mizuta2016investigation,meltonSIGecom}, and may be further incentivized to submit quantity in excess of their actual demand to increase the probability of receiving their desired allocation of a resource~\cite{farmer2012}.
\section{\sysname}
We now introduce \sysname, a reordering policy that 
alleviates the effect of uneven delays introduced by the market's infrastructure,
while remaining ``temporally fair'' at all times.

\subsection{Rationale}
Our premise is that there will always be participants who are 
advantaged in receiving updates and/or in submitting new orders, as
no distributed system can remain symmetrical and perfectly load balanced at all times. 
Instead, we seek to achieve temporal fairness by stochastically equalizing participants 
whose order submission times are within the error-margins measured by the exchange operator.

\subsection{Reordering Processes}
\sysname operates in two phases realised by Algorithms~\ref{alg:buffering} and~\ref{alg:draining} respectively.
Once a new order-message $o$ (structure details are available in Table~\ref{tab:fields}) arrives at the exchange,
the policy component intercepts it and executes Algorithm~\ref{alg:buffering}:
If the message is a Cancel order, it is then forwarded immediately to the matching
engine without any delay added. Otherwise, the message is placed in a buffer ($b$).
The buffer is determined based on the instrument the order seeks to capture, its side (buy or sell), its 
price and if it is ``marketable'' or not (i.e., crosses the spread or not).
A buy order at a price higher than at least one of the resting sell orders
is placed in the ``marketable'' buffer $M[instrument][buy]$ with a \textit{Null} price.
Similarly a sell order at a price lower than at least one of the resting buy orders
is placed in the ``marketable'' buffer $M[instrument][sell]$ with a \textit{Null} price.
Every other order, retains its price and is placed in a buffer specific to its 
price level i.e., $B[instrument][side][price]$.
The only exception are Immediate-or-Cancel (IOC) orders that are dropped if
they are not filled immediately.
Upon adding an order to a previously empty buffer, a ``drain'' timer starts counting
up to a predetermined amount of time.

Algorithm~\ref{alg:draining} is executed on $b$ once its predetermined timer
elapses.
Initially, it groups $b$'s messages into lists per participant ($P[\ ]$),
based on the identifier $o_{u}$ found in each order.
Subsequently, each participant's list is sorted and the orders
are ranked based on their arrival time. Orders who arrived earlier
are placed closer to the top of the list. Additionally, the 
algorithm generates a random ordering of the participant identifiers ($R$).
In its last stage, the algorithm iterates over the participants' 
lists (according to their random order in $R$).
On each iteration, the algorithm selects one of the lists, pops its top order
and forwards it to the matching engine of the exchange. The algorithm keeps 
iterating over the participants' lists until all of them are empty.\footnote{Alternative draining policies, i.e., approaches to `Algorithm 2' for Libra are disclosed in \cite{meltonFloor}.}

\begin{table}
	\centering	
	\resizebox{0.9\columnwidth}{!}{
	\begin{tabular}{c l l}
	\toprule
	\thead{Symbol} & \thead{Name} & \thead{Description}\\ \hline 
	\addlinespace
	\textit{$o_{u}$} & Identifier & Unique Market Participant Identifier \\
	\textit{$o_{i}$} & Instrument & Unique Identifier of asset traded \\
	\textit{$o_{t}$} & Type & Order types (e.g., Market, Limit, Cancel, IOC) \\
	\textit{$o_{s}$} & Side & Buy or Sell Order \\
	\textit{$o_{p}$} & Price & Limit order price \\
	\bottomrule 
	\addlinespace
	\end{tabular}
	}
	\caption{Each order $o$ has five fields that are considered by \sysname's reordering policy.}
	\label{tab:fields}
\end{table}

\begin{algorithm}
	\footnotesize
	\SetAlgoLined
	\SetKwInOut{Input}{Input}
	\Input{Order message $o$}

    $\textrm{instrument} \gets o_i$\Comment{Instrument traded}\\
    $\textrm{side} \gets o_s$\Comment{Buy or Sell?}\\
    $\textrm{type} \gets o_t$\Comment{Market, Limit, IOC}\\

	\If{$o_t = \textrm{Cancel}$}{
		$\textrm{Forward}(o)$\\
        $\textrm{Return()}$
    }\ElseIf{$\textrm{side} = \textrm{Buy and price} \geq \textrm{Best\_offer}$}{
        $b^* \gets \textrm{M[instrument][buy][Null]}$ \Comment{Reference to marketable ``buy'' buffer}\\
   	}\ElseIf{$\textrm{side} = \textrm{Sell and price} \leq \textrm{Best\_bid}$}{
        $b^* \gets \textrm{M[instrument][sell][Null]}$ \Comment{Reference to marketable ``sell'' buffer}\\
    }\ElseIf{$\textrm{type} \neq \textrm{IOC}$}{
        $\textrm{price} \gets o_p$\Comment{Limit price}\\
        $b^* \gets \textrm{B[instrument][side][price]}$ \Comment{Reference to price-specific buffer}\\
	}\Else{
	   $\textrm{Return()}$
	}
	
	$ $\\
	
	\If{$b^* \neq \emptyset$}{
        $b^*\textrm{.append(o)}$\\
	}\Else{
	   $b^* \gets \{o\}$\\
	   $\textrm{Start\_Timer}(b^*)$ \Comment{The 1st order starts the timer}\\
	}
\caption{\textbf{Buffering.} Collects each incoming order message $o$ and either forwards it to the matching engine directly or places it in its corresponding buffer. If the order is immediately marketable (i.e., crosses the spread) it is placed in the buy/sell buffer $M$, otherwise it is moved to a price-specific buffer from the set of buffers $B$.}
\label{alg:buffering}
\end{algorithm}

\begin{algorithm}
	\footnotesize
	\SetAlgoLined
	\SetKwInOut{Input}{Input}
	\Input{Buffer of order messages $b$, List of participant identifiers $U$}
	   
	 $P \gets \{\}$\\
	 \For {$o \in b$}{
	    $P[o_{u}] \gets P[o_{u}] \cup {o}$\Comment{List of orders per participant}\\
	 }
    \For {$u \in U$}{
	    $P[o_{u}] \gets \textrm{sort(}P[o_{u}])$ \Comment{Sort by ascending arrival time}\\
	  }
    
    $R \gets \textrm{shuffle(U)}$ \Comment{Random order of participant IDs}\\
    
    \While {$not (P[u] = \emptyset\ \forall u \in U$)}{
        \For {$u \in R$}{
            \If {$P[u] \neq \emptyset$}{
                $\textrm{Forward}(P[u].pop())$ \Comment{Pop the oldest order of $u$}\\
            }
        }
    }
\caption{\textbf{Draining.} The draining algorithm is executed whenever a timer (of a buffer $b$) goes off. It initially reorders the messages held by the buffer and then forwards them to the matching engine.}
\label{alg:draining}
\end{algorithm}

\subsection{System Properties}\label{subsec:secproperties}
In this Section, we examine if \sysname fulfills the 
fairness, robustness and minimum impact requirements we seek for modern markets.

\noindent\textbf{Correctness.}
Under ideal, non-faulty conditions, our specification substantially ensures that \sysname will handle orders
fairly at all times and ensures that slow participants will never be advantaged over faster ones.
However, participants with small differences in their response times may be ``equalized''
by our order-batching algorithm. As explained in Section~\ref{sec:fairness},
this does not violate our definition and it is even considered a 
desirable feature by some works (e.g., see ``social welfare'' considerations~\cite{farmer2012,harris2013,budish2015,budish2017will}
in Section~\ref{sec:related}).\\

\noindent\textbf{Robustness.}
We now look into various (failure and adversarial) scenarios,
examine the resilience of our policy and motive
the relevant design decisions. Our analysis here cannot
exhaustively present all possible attacks and failure scenarios.
Instead, we consider those that affect other reordering policies, 
as well as new ones that seek to exploit the specific characteristics 
of \sysname.

One of the attacks seeking to exploit the processes of \sysname 
is ``placeholding''. A fast participant ($P_1$) submits a ``placeholder''
order just before an economic event occurs so as to 1) intentionally trigger the 
buffer's timer early and 2) place their actual order shortly before the end of the delay period.
This attacks aims to effectively exclude disadvantaged participants ($P_2$) whose
reaction/transmission time is too long for them to submit an order before the batch closes.
For example, given an instrument trading at \$1.00 and a timer length of 3ms,
first $P_1$ bids at \$0.01, 2.9 milliseconds ahead of an economic news release.
This starts the batch timer for the instrument and enables $P_1$ to 
place its true bid or offer at 3.0ms (depending on the event's outcome)
while leaving $P_2$ who is slower, off the batch.
To avoid this attack, \sysname holds a separate race for each price level.
In this case, $P_1$ will trigger the timer for a race at a price-level
that $P_2$ will never compete in. 
Such attacks are also the reason that IOC orders
are not included in buffers and do not trigger any timers. If they were, 
participants could exploit them to start the timer early at no economic cost for them~\cite{meltonSIGecom}.

$P_1$ can also submit a placeholding order at the correct price-level but of the
minimum quantity possible (usually 10 or 100 shares in stock exchanges).
To disincentivize such practices, we use arrival-time sorting (lines 4-5 in Algorithm~\ref{alg:draining}).
In the presence of another comparably fast participant ($P_3$), who is not manipulative,
$P_1$'s actual order (submitted after the event) will always execute after the first order
of $P_3$, as the early ``placeholding'' order of minimum size will always be on top 
of $P_1$'s list (sorted by arrival time).

Moreover, unlike random-delays and FBA, \sysname is 
tolerant to manipulative participants who submit 
several duplicated order copies (see attack in Section~\ref{subsec:randomdelays}).
This resilience is due to the \textit{participant grouping} operation (Algorithm~\ref{alg:draining})
and the random iteration sequence that prevents participants from skewing their execution 
probabilities by adding more orders.
This mechanism can be gamed through proxy/Sybil accounts that 
operate under seemingly independent participant identities but are controlled by one firm only.  
In our live deployment of \sysname, we use information from Know-Your-Customer processes to detect and
merge parent and child companies that have common ownership interests.\\

\noindent\textbf{Minimum Impact.}
Order-triggered timers allow our policy 
to equalize participants without introducing delays 
that are long in relation to the magnitude of temporal unfairness.
In fact, \sysname achieves fairness 
even if the batch-length is equal to the magnitude of the 
temporal unfairness it seeks to address.
Let two competing participants $P_1$ and $P_2$,
whose order submission latency is $2ms$ and $3ms$ (due to uneven delays),
respectively. A \sysname deployment with delay $T\geq1ms$
guarantees that $P_1$ and $P_2$ will have equal chances of winning
in the race. Once $P_1$ submits their order, the timer starts and
$P_2$ will have at least 1ms to submit theirs.
As explained above, the participant shuffling ensures that
two orders that are in the same batch have the same probability
of being processed first.

Finally, the Cancel-order reordering exemption
(i.e., no delay added to Cancel orders)
is an \textit{optional} feature that aims to reduce
adverse selection for makers. It allows 
Cancel orders from slow market makers to outrace bids and offers 
from fast takers and thus prevents takers from engaging in Sniping 
and other practices that take advantage of mispriced bids from slow participants. 
While it may seem to be a fairness concern as it allows certain participants to overtake others,
it is common in financial markets that makers and takers are not subject to the same exact treatment~\cite{aldrich2019experiments,budish2015}.
\sysname can be also deployed without this feature but it has been observed on a different forex trading venue that treating makers and takers `the same' in a batching regime can actually exacerbate adverse selection/sniping~\cite{Golden2015}---a concern also expressed by others~\cite{lehalle2017limit}.

\section{Experimental Evaluation}\label{sec:evaluation}
In this Section, we evaluate \sysname experimentally by deploying it in
a live forex exchange ({\em Refinitiv Matching}~\cite{meltonSIGecom}) and monitoring various aspects of the market.
Our measurements cover the trading activity of 
a foreign exchange over a two-year trading period.
The exchange serves hundreds of trading firms
(i.e., participants) and thousands of users.
The combined market share of the exchange and its competing
platform is estimated to be 15\%~\cite{hasbrouck2017fx}.
The scales on some of the figures' axes have been redacted, as they contain  
commercially sensitive information. All the redacted scales 
cross the other axis at 0.

\subsection{Impact on Participants}
\begin{figure}
    \centering
    \includegraphics[width=\columnwidth]{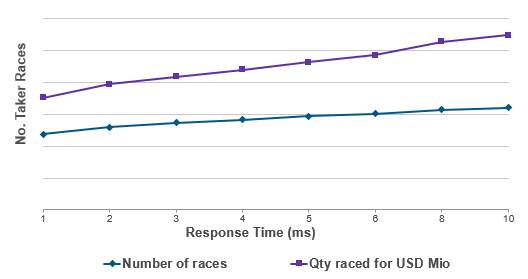}
    \caption{The effect of the batch length on the amount of ``taking'' that involves more than one participants. \sysname does not commensurately increase the amount of taking in contention, even when the latency is increased tenfold.}
    \label{fig:timer1}
\end{figure}

In this first experiment, we measure the impact that \sysname's latency (i.e., batch length)
has on participants on the venue. The results are shown in Figure~\ref{fig:timer1}.
Initially, we start from a batch length of $T=1ms$ which we gradually increase
up to $10ms$ in intervals of $1ms$. For each latency setting, we report the number of takers
races that took place (blue) and the quantity that competing participants raced for in USD Millions (purple).
As the latency increases, \sysname enables slower
participants to compete on equal footing with faster ones. Thus, the longer the  period, by definition,
the more likely it is that price taking will involve more than one participants.
However, as we see in Figure~\ref{fig:timer1} lengthening the batching period does not commensurately
increase the amount of taker races (i.e., instances in which more than one distinct participant is racing to hit the same bid or offer). In fact, a tenfold increase in the 
the batching period (from 1ms to 10ms) resulted in a less than 50\% increase in both the number of races and the quantity raced for. 
What this seems to indicate is that there are diminishing returns from what some have argued involves disadvantageously imposing longer (rather than shorter) delays in order processing on a venue (see \cite{baldauf2018fast,brolley2018order}).

\begin{figure}
    \centering
    \includegraphics[width=\columnwidth]{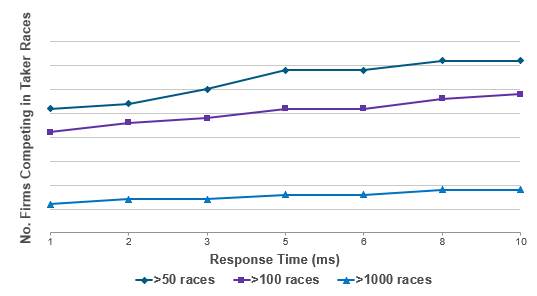}
    \caption{Increases in latency seem to affect primarily participants who compete in more than 1000 races per week, while participants who occasionally compete by coincidence seem less affected.}
    \label{fig:timer2}
\end{figure}

In Figure~\ref{fig:timer2}, we stratified our dataset based on the number of taker races each participant (trading firm) competed in weekly.
This illustration aims to show the number of firms that are affected by the increase in latency imposed by the setting the length of the batch to increasing values. 
We plot three main classes: Firms that competed weekly in $>$50 races (green-diamond), 
in $>$100 races (purple-rectangle), and in $>$1000 races (blue-triangle). The
first two classes ($>$50 and $>$100 races) likely include participants who 
compete with others by coincidence, whereas the third one ($>$1000) shows  participants who compete habitually. 
We observe that the tenfold increase in latency (1ms to 10ms) less than doubled the number of firms competing (class 3), while it had an even  smaller effect (${\sim}25\%$) on classes 1 and 2.
We also observe that most participants do not compete with others frequently and
only a small segment of the market seem to be using such strategies. 
These differences in latency are relevant only to fast and sophisticated traders, 
as investors who are not automated and in physical proximity cannot deploy strategies at these 
timescales. This again may suggest that shorter delays imposed for the purposes of batching are preferable to longer ones.

\subsection{Competing vs. Non-Competing Participants}
\begin{figure}
\centering
\includegraphics[width=\columnwidth]{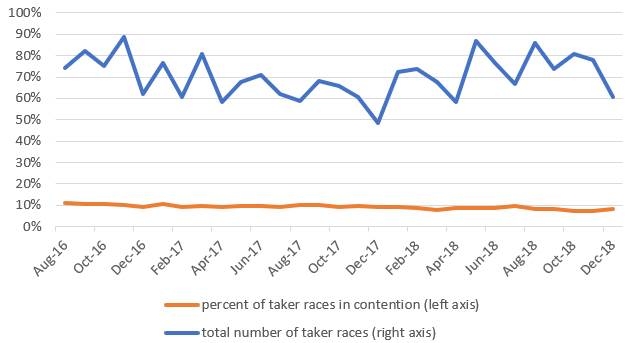}
\caption{Despite the fluctuations in the total amount of price taking, the proportion of price taking 
that is in contention remained stable (${\sim}10\%$) since the deployment of \sysname.}
\label{fig:contention}
\end{figure}

To better understand the degree that \sysname affects 
participants that directly compete with others, we measured the total 
number of instances where participants tried to take a price from the market (blue)
and the percentage of races where two or more participants were competing (orange).
Figure~\ref{fig:contention} shows the monthly trend of these two variables since the
deployment of \sysname.
We observe that while the total amount of taking fluctuates with the trading volumes per month, 
the level of price taking that is in contention remains at approximately 10\% throughout the whole period measured. 
What this seems to indicate is that the deployment of \sysname did not cause a significant change in the amount of price-taking activity that is in contention. 

\subsection{Sniping Protection}
\begin{figure}
    \centering
    \includegraphics[width=\columnwidth]{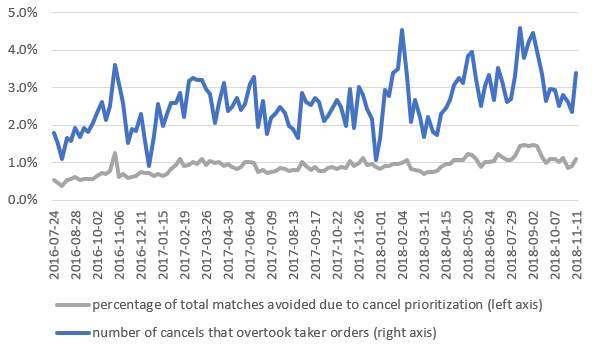}
    \caption{Adverse selection graph that illustrates the effectiveness of \sysname's cancel-order prioritization feature in preventing sniping.}
    \label{fig:selection}
\end{figure}

As discussed in the previous section, Cancel-order prioritization is an optional feature of
\sysname that aims to prevent adverse selection against market makers. Figure~\ref{fig:selection}
shows the number of cancel orders that overtook other orders from takers (blue).
Additionally, it illustrates the percentage of matches that were prevented due to a cancel order
overtaking a taker's ``sniping'' order (grey). 
While the number of ``overtakes'' fluctuates a lot over time, the percentage of matches avoided due
to the prioritization of cancels steadily grows from 0.5\% to over 1\%.
This highlights that the (optional) exemption feature not only prevented a common side-effect of reordering policies (i.e., increased sniping~\cite{lehalle2017limit}) but also positively contributed in mitigating the phenomenon in general. 

\section{Related Work}\label{sec:related}
Our discussion of order-matching policies in this work has centered on
our strict definition of temporal fairness and the requirements that 
market designs need to fulfill in order to be practically applicable.\\

\noindent\textit{Social welfare}\\
Besides these topics, the market design literature~\cite{farmer2012,harris2013,budish2015,budish2017will} 
has also explored the notion of social welfare pertaining to market participant's speeds.
The argument that appears throughout that literature is (roughly) that the technology `arms race' in pursuit of speed is socially wasteful, a form of the prisoner's dilemma, and all market participants would benefit
if all electronic financial markets were re-designed to de-emphasize the effects of participant speed on resource allocation. 
As compelling as this social value argument for
``slowing'' markets down might be, it is far from settled
as to whether doing so actually improves them. In particular,
several works have argued that such interventions have the 
opposite effect than the intended~\cite{spatt2018,chen2017value,brown2016slowing,FIAPTG,duffin2018agent,brolley2018order,aoyagi2019strategic} \cite[fn.8]{melton2018fairware}. For example, pertaining to the regulatory approval of 
the Intercontinental Exchange's speedbump (i.e., 3ms constant delay)~\cite{icespeedbump2019},
the objecting commissioners argued that:
``Risk (and reward) move at the speed of information. Those that invent, and invest in, faster information transmission technologies to capitalize on market dislocations reap the profits of their advantage.''~\cite{statement2019}.
Overall, the critics of speed de-emphasization argue that the competition between traders drives technological advances and 
enhances the accessibility and responsiveness of the markets.\\

\noindent\textit{Technical Market Manipulation}\\
Another line of research has recently focused on a particular set of technical problems, 
that are also related to ``speed'' in continuous electronic markets. These works study unfairness from the perspective of systems security and approaches 
policy gaming strategies as security problems or problems of {\em technical} (cf. economic) market manipulation~\cite{spatt2014security,Mavroudis19,aitken2015high}. 
In this category of problems, market participants exploit technical flaws in the electronic trading venue's implementation to gain speed advantages over other participants. Examples of such attacks and exploitable vulnerabilities have appeared in various patent applications~\cite{CMEOptimistic,simonoff2017}, disciplinary notices~\cite{CME17}, presentations~\cite{CME14}, and a forthcoming book~\cite{Hurd}.

\section{Conclusions}
This work aims to bridge the gap between theoretical market models
and their real-life implementations in electronic markets.
Towards this goal, we provided a novel definition of ``temporal fairness'' and a set of requirements 
that enable researchers and practitioners to realistically evaluate market designs and their 
robustness to technical inefficiencies. 
Based on these, we then analysed several existing reordering policies and
used our observations to design a novel semi-continuous market that de-emphasizes 
speed advantages and retains its fairness properties regardless of infrastructure 
inefficiencies and gaming attempts. Our experimental results seem to support our initial 
analysis and show that \sysname reduces speed-based predatory trading without severely 
impacting existing trading strategies.
\begin{acks}
We would like to thank the anonymous reviewers for their valuable feedback.
We would also like to thank George Danezis for the helpful discussions, 
as well as Lara, our research assistant. Vasilios Mavroudis was supported
by Binance X through the Tradescope project.
\end{acks}

\bibliographystyle{ACM-Reference-Format}
\bibliography{bibliography}
\end{document}